\documentclass[twocolumn,showpacs,preprintnumbers,amsmath,amssymb]{revtex4} 

\begin{document}

\title{A Zoology of Bell inequalities resistant to detector inefficiency}
\author{Serge Massar$^{1,2}$, Stefano Pironio$^{1}$ J\'er\'emie Roland$^{2}$ 
and Bernard Gisin$^{3}$}

\address{$^1$ Service de Physique Th\'eorique, CP 225,
Universit\'e Libre de Bruxelles, 1050 Brussels, Belgium \\
$^{2}$ Ecole Polytechnique, CP 165, Universit\'e Libre de Bruxelles, 1050
Brussels, Belgium\\
$^{3}$ Group of Applied Physics, University of Geneva, 20 rue de 
l'Ecole-de-M\'edecine, CH-1211 Geneva 4, Switzerland} 
\date{\today}
\begin{abstract}
We derive both numerically and analytically Bell inequalities and quantum 
measurements that present enhanced resistance to detector inefficiency. In 
particular we describe several Bell inequalities which appear to be optimal with 
respect to inefficient detectors for small dimensionality $d=2,\: 3, \: 4$ and
2 or more measurement settings at each side. We also generalize the family of
Bell inequalities described in Collins et all \cite{collins} to take into 
account the inefficiency of detectors. In addition we consider the 
possibility for pairs of entangled particles to be produced with probability 
less than one. We show that when the pair production probability is small, one 
must in general use different Bell inequalities than when the pair production 
probability is high. \end{abstract}
\maketitle

\section{Introduction}

A striking feature of quantum entanglement is
non-locality. Indeed, as first shown by Bell in 1964 \cite{bell} classical local
theories cannot reproduce all the correlations exhibited by entangled quantum
systems. This non-local character of entangled states is demonstrated in
EPR experiments through the violation of Bell inequalities. However due to
experimental imperfections and technological limitations, Bell tests suffer from
loopholes which allow, in principle, the experimental data to be reproduced
by a local realistic description. The most famous of these loopholes are the
locality loophole and the detection loophole. Experiments carried on photons
have closed the locality loophole \cite{weihs} and recently Rowe et al closed
the detection loophole using trapped ions \cite{rowe}. But so far, 30 years
since the first experiments, both loopholes have not been closed in a
\emph{single} experiment.

The purpose of this paper is to study how one can devise new tests of
non-locality able to lower the detector efficiency necessary to
reject any local realistic hypothesis. This could be a way towards a 
loophole-free test of Bell inequalities and is important for several 
reasons. First, as quantum entanglement is the basic ingredient of quantum 
information processing, it is highly desirable to possess undisputable tests of 
its properties such as non-locality. Even if one is convinced (as we almost all 
are) that nature is quantum mechanical, we can imagine practical situations 
where it would be necessary to perform loophole-free tests of Bell inequalities. 
For example, suppose you buy a quantum cryptographic device based on Ekert 
protocol. The security of your cryptographic apparatus relies on the fact that 
you can violate Bell inequalities with it. But if the detectors efficiencies 
aren't sufficiently high, the salesman can exploit it and sell to you a 
classical device that will mimic a quantum device but which will enable him to 
read all your correspondence. Other reasons to study the resistance of quantum 
tests to detector inefficiencies are connected to the classification of 
entanglement. Indeed an important classification of entanglement is related to 
quantum non-locality. One proposed criterion to gauge how much non-locality is 
exhibited by the quantum correlations is the resistance to noise. This is what 
motivated the series of works \cite{zeilig, durt} that led to the generalization 
of the CHSH inequality to higher dimensional systems \cite{collins}. The 
resistance to inefficient detectors is a second and different criterion that we 
analyse in this paper. It is closely related to the amount of classical 
communication required to simulate the quantum correlations \cite{massar}.

The idea behind the detection loophole is that in the presence of unperfect
detectors, local hidden variables can "mask" results in contradiction with
quantum mechanics by telling the detectors not to fire. This is at the origin of
several local hidden variable models able to reproduce particular quantum
correlations if the detector efficiencies are below some threshold value
$\eta_*$ (see \cite{gisin, lhv,inprep} for example). In this paper, we introduce 
two parameters that determine whether a detector will fire or not: $\eta$, the 
efficiency of the detector and $\lambda$, the probability that the pair of 
particles is produced by the source of entangled systems. This last parameter 
may be important for instance for sources involving parametric down conversion 
where $\lambda$ is typically less than $10\%$. So far, discussions on the 
detection loophole where concentrating on $\eta$, overlooking $\lambda$. However 
we will show below that both quantities play a role in the detection loophole 
and clarify the relation between these two parameters. In particular we will 
introduce two different detector thresholds: $\eta_*^{\lambda}$, the value above 
which quantum correlations exhibit non-locality for given $\lambda$, and 
$\eta_*^{\forall \lambda}$, the value above which quantum correlations exibit 
non-locality for any $\lambda$.

We have written a numerical algorithm to determine these two thresholds for given
quantum state and quantum measurements. We then searched for optimal measurements
such that $\eta_*^{\lambda=1}$ and $\eta_*^{\forall \lambda}$ acquire the lowest
possible value. In the case of bipartite two dimensional systems the most
important test of non-locality is the CHSH inequality \cite{chsh}.
Quantum mechanics violates it if the detector efficiency $\eta$ is above
$=2/(\sqrt{2}+1) \approx 0.8284$ for the maximally entangled state of two
qubits. In the limit of large dimensional systems and large number of settings, 
it is shown in \cite{massar} that the efficiency threshold can be arbitrarily 
lowered. This suggests that the way to devise optimal tests with respect to the 
resistance to detector inefficiencies is to increase the dimension of the 
quantum systems and the number of differents measurements performed by each 
party on these systems. (This argument will be presented in more details in 
\cite{inprep}). We have thus performed numerical searches for increasing 
dimensions and number of settings starting from the two qubit, two settings 
situation of the CHSH inequality. Our results concern "multiport beam 
splitters measurements" \cite{multiport} performed on maximally entangled 
states. They are summarized in Table I. Part of these results are accounted for 
by existing Bell inequalities, the other part led us to introduce new Bell 
inequalities. \\

\begin{table} 
\label{tab:table}
\caption{
Optimal threshold detector efficiency for
varying dimension $d$ and number of settings ($N_a \times N_b$) for the
detectors. $\eta_*^{\lambda=1}$ is the threshold efficiency for a source such 
that the pair production probability $\lambda=1$ while $\eta_*^{\forall\lambda}$ 
is the threshold efficiency independent of $\lambda$. The column $p$ gives the 
amount of white noise $p$ that can be added to the entangled state so that it 
still violates locality (we use for $p$ the same definition as that given in 
\cite{zeilig, durt}). The last column refers to the Bell inequality that 
reproduce the detection threshold. Completely new inequalities introduced in 
this paper are indicated by "New".} \begin{ruledtabular} 
\begin{tabular}{lcccccc} $d$ & $N_a\times N_b$ &$\eta_{*}^{\lambda=1}$ 
&$\eta_{*}^{\forall\lambda}$ & $p$ &Bell inequality  \\ \hline 2 & $2\times 2$ & 
$0.8284$ & $0.8284$ & 0.2929 & CHSH   \\ 2 & $3\times 3$ & $0.8165$ & & 0.2000 
&\begin{tabular}{c}  New\\ \footnotesize(see also ref \cite{bell,wigner}) 
\end{tabular}\\ 2 & $3\times 3$ &  & $0.8217$ & 0.2859 & New \\ 2 & $3\times 4$ 
& & $0.8216$ & 0.2862 & New \\ 2 & $4\times 4$ &  & $0.8214$ & 0.2863 &New   \\ 
\hline 3 & $2\times 2$ & $0.8209$ & $0.8209$ &0.3038 &based on 
ref.\cite{collins}\\ 3 & $2\times 3$ & $0.8182$ & $0.8182$ &0.2500 & 
\begin{tabular}{c}  New\\ \footnotesize(related to ref \cite{helle}) 
\end{tabular} \\ 3 & $3\times 3$ & $0.8079$ &  &0.2101 & New  \\ 3 & $3\times 3$ 
&  & $0.8146$ &0.2971 & New  \\ \hline 4 & $2\times 2$ & $0.8170$ & $0.8170$ & 
0.3095 & based on ref.\cite{collins}\\ 4 & $2\times 3$ & & $0.8093$ &  0.2756& 
New \\ 4 & $3\times 3$ & & $0.7939$ &  0.2625 & New \\ \hline 5 & $2\times 2$ & 
$0.8146$ & $0.8146$ & 0.3128 & based on ref.\cite{collins}\\ \hline 6 & $2\times 
2$ & $0.8130$ & $0.8130$ & 0.3151 &based on ref.\cite{collins}\\ \hline 7 & 
$2\times 2$ & $0.8119$ & $0.8119$ & 0.3167 &based on ref.\cite{collins}\\ \hline  
$\infty$ & 2 $\times$ 2 & 0.8049 & 0.8049 &0.3266 &based on ref.\cite{collins} 
\\ \end{tabular} \end{ruledtabular} \end{table} 

The main conclusions that can be drawn from this work are:
\begin{enumerate}
\item Even in dimension 2, one can improve the resistance to inefficient 
detectors by increasing the number of settings. 
\item One can further increase the resistance to detection inefficiencies by increasing the dimension. 
\item There are different optimal measurements settings and Bell 
inequalities for a source that produces entangled particles with high probability ($\lambda\approx 1$) and one 
that produces them extremely rarely ($\lambda \rightarrow 0$). Bell inequalities associated with this last situation
provide a detection threshold that doesn't depend on the value of the pair production probability.
\item For the measurement scenarios numerically 
accessible, only small improvements in threshold detector efficiency are 
achieved. For instance the maximum change in threshold detector
efficiency we found is approximatively $4\%$
\end{enumerate}

The paper is organized as follows:
First, we review briefly the principle of an EPR experiment in section
\ref{secQC} and under which condition such an experiment admits a
local-realistic description in section \ref{secLHV}. In section \ref{secDE} we
clarify the role played by $\eta$ and $\lambda$ in the detection loophole. We
then present the technique we used to perform the numerical searches in
\ref{secNS} and to construct the new Bell inequalities presented in this
paper in \ref{secBI}. Section \ref{secRE} contains our results. In particular in 
\ref{secRE2x2} we generalize the family of inequalities introduced in 
\cite{collins} to take into account detection inefficiencies and in 
\ref{secRE3x3} we present the two different Bell inequalities associated to the 
two-dimensional three by three settings measurement scenario. In the Appendix, 
we collect all the measurement settings and Bell inequalities we have obtained.

\section{General Formalism}

\subsection{Quantum correlations}\label{secQC}

Let us  review the principle of an a EPR experiment: two parties,
Alice and Bob, share an entangled state
$\rho_{AB}$. We take each particle to  belong to a $d$ dimensional Hilbert
space. The parties carry out measurements on their particles.
Alice can
choose between $N_a$ different von Neumann measurements
$A_i$ ($i=1,\ldots,
N_a$)
and Bob can choose between $N_b$ von Neumann  measurements
$B_j$ ($j=1,\ldots, N_b$).
Let $k$ and
$l$ be Alice's and Bob's outcomes. We
suppose that the
number of possible outcomes is the same for each party and that the
values of $k$ and $l$ belong to $\{ 0,\ldots,d-1\}$. To each
measurement $A_i$ is thus associated a complete set of $d$ orthogonal
projectors
$A_i^k = |A_i^k\rangle\langle A_i^k|$
and similarly for $B_j$.
Quantum mechanics predicts the following probabilities for the outcomes
\begin{eqnarray}
\label{qcid}
P^{QM}_{kl}(A_i,B_j)&=&\mbox{Tr}((A_i^k\otimes B_j^l) \rho_{ab})\ , \nonumber \\
P^{QM}_l(B_j)&=&\mbox{Tr}((\openone_A\otimes B_j^l)\rho_{ab})\ ,\nonumber \\
P^{QM}_k(A_i)&=&\mbox{Tr}((A_i^k\otimes \openone_B)\rho_{ab})\ . \nonumber \\
\end{eqnarray}

In a real
experiment, it can happen that the measurement gives no outcome,
due to detector
inefficiencies, losses or because the pair of entangled states has not
been produced. To take into account these cases in the most general
way,
we enlarge the
space of possible outcomes and add a new
outcome, the ``no-result outcome", which we label $\emptyset$.  Quantum
mechanics now predicts a modified set of correlations:
\begin{eqnarray}
P^{QM}_{\lambda \eta}(A_i=k,B_j=l)&=&
\lambda\eta^2P^{QM}_{kl}(A_i,B_j) \quad k,l\neq
\emptyset \ ,\nonumber \\
P^{QM}_{\lambda \eta}(A_i=\emptyset,B_j=l)&=&
\lambda\eta(1-\eta)P^{QM}_l(B_j) \quad l \neq
\emptyset \ ,
\nonumber \\
P^{QM}_{\lambda \eta}(A_i=k,B_j=\emptyset)&=&
\lambda\eta(1-\eta)P^{QM}_k(A_i) \nonumber
\quad k \neq \emptyset \ ,\\
P^{QM}_{\lambda \eta}(A_i=\emptyset,B_j=\emptyset)&=&
1-\lambda+\lambda(1-\eta)^2 \ . \nonumber \\ \label{qc}
\end{eqnarray}
where $\eta$ is the detector efficiency, and
$\lambda$ is the probability that a pair of particles is produced
by the source of entangled systems. By detection efficiency $\eta$
we mean the probability that the detector gives a result if a particle was
produced, i.e. $\eta$ includes not only the "true" efficiency of the
detector but also all possible losses of the particle on the path from the source
to the detectors.

\subsection{Local Hidden Variable Theories \& Bell Inequalities}\label{secLHV}

Let us now define when the results (\ref{qc}) of an EPR experiment can
be explained by a local hidden variable (lhv) theory. In a
lhv theory, the outcome of Alice's measurement is determined by the setting
$A_i$ of Alice's measurement apparatus and by a random variable shared by both
particles. This result should not depend on the setting of Bob's measurement
apparatus if the measurements are carried out at spatially separated
locations. The situation is similar for Bob's outcome. We can describe without
loss of generality such a local variable theory by a set of $(d+1)^{N_a+N_b}$
probabilities $p_{K_1\ldots K_{N_a}L_1\ldots L_{N_b}}$ where Alice's local variables
$K_i \in\{0,\ldots,d-1, \emptyset\}$ specifies the result of measurement $A_i$ and
Bob's variables $L_j\in \{0,\ldots,d-1, \emptyset\}$ specify the result of
measurement $B_j$. The correlations $P(A_i=K,B_k=L)$ are
obtained from these joint probabilities as marginals. The
 quantum predictions
can then be reproduced by a lhv theory if and only if the following
 $N_aN_b(d+1)^2$
equations are obeyed:
\begin{eqnarray}
\label{clprob}
\sum_{\bf K  L} p_{\bf K  L}
\delta_{K_i,K}\delta_{L_j,L}\!\!&=&\!\!P^{QM}_{\lambda
\eta}(A_i=K,B_j=L) \end{eqnarray}
with the conditions:
\begin{eqnarray} \label{cond1}
\sum_{\bf K  L} p_{\bf K L } &=&1\ ,\\
p_{\bf K L} \geq 0 \ , \label{poscond}
\end{eqnarray}
where we have introduced the notation
${\bf K} = K_1\ldots K_{N_a}$ and ${\bf L} = L_1\ldots L_{N_b}$.
Note that the equations (\ref{clprob}) are not all independent since
quantum and classical probabilities share additional constraints such as the
normalization conditions:
\begin{equation}
\label{normalization}
\sum_{K,L} P(A_i=K,B_j=L)=1
\end{equation}
or the no-signalling conditions:
\begin{equation}
\label{nosignalling}
P(A_i=K)=\sum_L P(A_i=K,B_j=L)\quad\forall\ j
\end{equation}
and similarly for $B_j$.

An essential result is that the
necessary and sufficient conditions for a given probability distribution
$P^{QM}$ to be reproducible by a lhv theory can be expressed, alternatively to
the equations (\ref{clprob}), as a set of linear inequalities for $P^{QM}$, the 
Bell inequalities. They can be written as 
\begin{equation}
\label{ineq}
I=I_{rr}+I_{\emptyset r}+I_{r \emptyset}+I_{\emptyset \emptyset}\leq c
\end{equation}
where
\begin{eqnarray} \label{belexpr}
I_{rr}&=&\sum_{i,j}\sum_{k,l \neq \emptyset} c_{ij}^{kl}P(A_i=k,B_j=l)
\nonumber \\
I_{\emptyset r}&=&\sum_{i,j}\sum_{l \neq \emptyset} c_{ij}^{\emptyset
l}P(A_i=\emptyset,B_j=l) \nonumber \\
I_{r\emptyset }&=&\sum_{i,j}\sum_{k \neq \emptyset}
c_{ij}^{k\emptyset}P(A_i=k,B_j=\emptyset)\nonumber \\
I_{\emptyset\emptyset}&=&\sum_{i,j}
c_{ij}^{\emptyset \emptyset}P(A_i=\emptyset,B_j=\emptyset). \nonumber \\ \end{eqnarray}
For certain values of $\eta$ and $\lambda$, quantum mechanics can violate one of 
the Bell inequalities (\ref{ineq}) of the set. Such a violation is the signal 
for experimental demonstration of quantum non-locality.

\subsection{Detector efficiency \& pair production probability}\label{secDE}

For a given quantum mechanical probability distribution $P^{QM}$ and
given pair production probability $\lambda$, the
maximum value of the detector efficiency $\eta$ for which there exists
a lhv variable model will be denoted $\eta_{*}^\lambda(P^{QM})$.
It has been argued \cite{gisin, lambda} that $\eta_{*}$ should not depend on 
$\lambda$. The idea behind this argument is that the outcomes $(\emptyset, 
\emptyset)$ obtained when the pair of particles is not created are trivial and 
hence it seems safe to discard them. A more practical reason, is that the pair
production rate is rarely measurable in experiments. Whatever, the logical
possibility exists that the lhv theory can exploit the pair production rate.
Indeed, we will show below that this is  the case when the number of settings of
the measurement apparatus is larger than 2. This motivates our definition of
threshold detection efficiency valid for all values of $\lambda$
\begin{equation} 
\eta_{*}^{\forall \lambda} = \max_{\lambda \neq
0}(\eta_{*}^{\lambda})=\lim_{\lambda \rightarrow 0} \eta_*^{\lambda} 
\end{equation}
The second equality follows from the fact that if a lhv model exists for a given 
value of $\lambda$ it also exists for a lower value of $\lambda$.
 
Let us study now the structure of the Bell expression $I(QM)$ given by 
quantum mechanics. This will allow us to derive an expression for 
$\eta_*^{\forall \lambda}$. Inserting the quantum probabilities (\ref{qc}) into 
the Bell expression of eq. (\ref{ineq}) we obtain 
\begin{multline} \label{qmineq} 
I(QM)=\lambda\eta^2 I_{rr}(QM) + \lambda\eta(1-\eta) I_{\emptyset r}(QM)\\
+ \lambda\eta(1-\eta)I_{r \emptyset}(QM)+(1 
+\lambda(\eta^2-2\eta))\sum_{i,j}c_{ij}^{\emptyset \emptyset} 
\end{multline}
where $I^{QM}_{rr}$ is obtained by replacing
$P(A_i=k,B_j=l)$ with $P^{QM}_{kl}(A_i,B_j)$ in $I_{rr}$
and $I^{QM}_{\emptyset r}$ by replacing $P(A_i=\emptyset, B_j=l)$ with 
$P^{QM}_l(B_j)$ in $I_{\emptyset r}$ and similarly for $I^{QM}_{r 
\emptyset}$.

For $\eta=0$, we know there exists a trivial lhv model and so the Bell
inequalities cannot be violated. Replacing $\eta$ by 0 in (\ref{qmineq}) we 
therefore deduce that 
\begin{equation}
\sum_{i,j}c_{ij}^{\emptyset \emptyset} \leq c.
\end{equation}
This divides the set of Bell inequalities into two groups: those such that
$\sum_{i,j}c_{ij}^{\emptyset \emptyset} < c$ and those for which $\sum_{i,j}c_{ij}^{\emptyset
\emptyset}=c$. Let us consider the first group. For small $\lambda$, these 
inequalities will cease to be violated. Indeed, take $\eta=1$ (which is the 
maximum possible value of the detector efficiency), then (\ref{qmineq}) reads 
\begin{equation}
I(QM)=\lambda I^{QM}_{rr}+(1-\lambda)\sum_{i,j}c_{ij}^{\emptyset \emptyset}. 
\end{equation}
The condition for violation of the Bell inequality is 
$I(QM)>c$. But since $\sum_{i,j}c_{ij}^{\emptyset\emptyset} <c$, for sufficiently 
small $\lambda$ we will have $I(QM)<c$ and the inequality will not be 
violated. These inequalities can therefore not be used to derive threshold   
$\eta_*^{\forall \lambda}$ that do not depend on $\lambda$, but they are still 
interesting and will provide a threshold $\eta_*^{\lambda}$ depending on 
$\lambda$.  Let us now consider the inequalities such that 
$\sum_{i,j}c_{ij}^{\emptyset \emptyset} =c$. Then $\lambda$ cancels in (\ref{qmineq}) and 
the condition for violation of the Bell inequality is that $\eta$ must be 
greater than \begin{equation}\label{etathr} \eta_{*}^{\forall \lambda} (P^{QM}) 
= \frac{ 2 c -  I^{QM}_{\emptyset r} - I^{QM}_{r \emptyset}}{c + I^{QM}_{rr} -
I^{QM}_{\emptyset r} - I^{QM}_{r \emptyset}} .
\end{equation}

It is interesting to note that if quantum mechanics violates a Bell 
inequality for perfect sources $\lambda=1$ and perfect detectors $\eta=1$, 
then there exists a Bell inequality that will be violated for $\eta<1$ and 
$\lambda\rightarrow 0$. That is there necessarily exists a Bell inequality 
that is insensitive to the pair production probability. Indeed the violation of 
a Bell inequality in the case $\lambda=1$, $\eta=1$ implies that there exists a 
Bell expression $I_{rr}$ such that $I_{rr}(QM)>c$ with $c$ the maximum value 
of $I_{rr}$ allowed by lhv theories. Then let us build the following inequality 
\begin{equation}\label{ibuild}
I=I_{rr}+I_{r \emptyset}+I_{\emptyset 
r}+\sum_{i,j}c_{ij}^{\emptyset \emptyset} P(A_i=\emptyset,B_j=\emptyset) \leq c
\end{equation}
where $\sum_{i,j}c_{ij}^{\emptyset \emptyset} =c$ 
and we take in $I_{r \emptyset}$ and $I_{\emptyset r}$ sufficiently negative 
terms to insure that $I \leq c$.  For this inequality, $\eta_*^{\forall 
\lambda}=(2c-I^{QM}_{\emptyset r}-I^{QM}_{r 
\emptyset})/(c+I^{QM}_{rr}-I^{QM}_{\emptyset r}-I^{QM}_{r \emptyset})<1$, which 
shows that Bell inequalities valid $\forall \lambda$ always exist. One can, in 
principle, optimize this inequality by taking $I_{r \emptyset}$ and $I_{\emptyset r}$ as 
large as possible while ensuring that (\ref{ibuild}) is obeyed.

From the experimentalist's point of view, Bell tests involving 
inequalities that depend on $\lambda$ need all events to be taken into 
account, including ($\emptyset$, $\emptyset$) outcomes, while in tests 
involving inequalities insensitive to the pair production  probability, it is 
sufficient to take into account events where at least one of the parties 
produces a result, i.e. double non-detection events ($\emptyset, \emptyset)$ can 
be discarded. Indeed, first note that one can always use the normalization 
conditions (\ref{normalization}) to rewrite a Bell inequality such as 
(\ref{ineq}) in a form where the term $I_{\emptyset \emptyset}$ does not appear. 
Second, if the events $(\emptyset,\emptyset)$ are not recorded in an experiment, 
the measured probabilities are relative frenquencies computed on the set of all 
events involving at least one result on one side. The probabilities measured in 
such experiments can be obtained from the probabilities (\ref{qc}) by replacing 
$\lambda$ with $\lambda'=\lambda/(1-(1-\eta)^2)$. While this rescaling of 
$\lambda$ is legitimate for inequalities that do not depend on the value of 
$\lambda$, it is however incorrect to perfom it for inequalities depending of 
$\lambda$, in particular this will affect the detection threshold.

\subsection{Numerical search}\label{secNS}

We have carried numerical  searches to find measurements such that the
thresholds $\eta_{*}^{\lambda=1}$ and $\eta_{*}^{\forall \lambda}$ acquire
the lowest possible value. This search is carried out in two
steps. First of all, for given quantum mechanical probabilities, we
have determined the maximum value of $\eta$ for which there exists a
local hidden variable theory. Second we have searched over the
possible measurements to find the minimum values of $\eta_{*}$.

In order to carry out the first step,
we have used the fact that the question of whether there
are classical joint probabilities that satisfy
(\ref{clprob}) with the conditions (\ref{cond1},\ref{poscond})
 is a typical linear optimization problem for which there exist
efficient algorithms \cite{zuk}. We have written a program which, given
$\lambda$, $\eta$ and a set of quantum measurements,  determines
whether
(\ref{clprob}) admits
a solution or not. $\eta_{*}^{\lambda}$ is then determined by performing a
dichotomic search on the maximal value of $\eta$ so that the set of constraints
is satisfied.

However when searching for
$\eta_{*}^{\forall \lambda}$ it is possible to dispense with the
dichotomic search by  using the following trick.
First of all because all the equations in eq. (\ref{clprob}) are not
independent, we can remove the constraints which involve on the right
hand side the probabilities
$P(A_i=\emptyset,B_j=\emptyset)$. Second we define rescaled variables
$\lambda  (1 - (1-\eta)^2) \tilde p_{\bf K L} =  p_{\bf K L}$.
Inserting the quantum probabilities eq. (\ref{qc})
we obtain the set of equations
\begin{eqnarray}
\sum_{\bf K L}
\tilde p_{\bf K L} \delta_{K_i,k}\delta_{L_j,l}\!\!&=&\!\! \alpha P^{QM}_{kl}
(A_i,B_j) \quad k,l \neq \emptyset\nonumber
\\
\sum_{\bf K L}\tilde p_{\bf K L}
\delta_{K_i,\emptyset}\delta_{L_j,l}\!\!&=&\!\! (1 -{\alpha\over 2}) P^{QM}_{l}
(B_j) \quad l \neq \emptyset\nonumber \\
\sum_{\bf K L}
\tilde p_{\bf K L}
\delta_{K_i,k}\delta_{L_j,\emptyset}\!\!&=&\!\! (1 -{\alpha\over 2}) P^{QM}_{k}
(A_i) \quad k \neq \emptyset\nonumber\\
\label{eqqq}
\end{eqnarray}
with the normalization
\begin{eqnarray}
\sum_{\bf K L}
\tilde p_{\bf K L}
 &=&  { 1 \over \lambda} {1 \over 1 - (1-\eta)^2}\label{eqqq4}
\end{eqnarray}
where $\alpha = \eta^2 / (1-(1-\eta)^2)$.
Note that $\lambda$ only appears in the last equation. We
want to find the maximum $\alpha$ such that these equations are obeyed
for all $\lambda$. Since $0< \lambda \leq 1$ \footnote{Actually
(\ref{weakcond}) corresponds to $0<\lambda\leq \frac{1}{1-(1-\eta)^2}$ so that
$\lambda$ can be greater than 1. But as stated earlier,  if a lhv model 
exists for a given value of $\lambda$ it is trivial to extend it to a lhv model 
for a lower value of $\lambda$. The maximum of $\eta_*^{\lambda}$ over the set 
$\lambda \in \: ]0,1/(1-(1-\eta)^2)]$ will thus be equal to the maximum over the 
set $\lambda \in \: ]0,1]$.}, we  can replace the last equation by the  
condition \begin{equation}\label{weakcond}
\sum_{\bf K L} \tilde p_{\bf K L}\geq 1.
\end{equation}
We thus are led to
search for the maximum $\alpha$ such that eqs. (\ref{eqqq}) are satisfied and
that the $\tilde p_{\bf K L}$ are positive and obey condition (\ref{weakcond}).
In this form the search for $\eta_{*}^{\forall \lambda}$ has become a linear
optimization problem and can be efficiently solved numerically.

Given the two algorithms that compute
$\eta_{*}^{\lambda=1}$ and $\eta_{*}^{\forall \lambda}$ for given settings,
the last part of the program is to find the optimal measurements.
In our search over the space of quantum strategies we first considered
the maximally entangled state
$\Psi=\sum_{m=0}^{d-1}|m\rangle_a |m\rangle_b$ in dimension $d$. The
possible
measurements $A_i$ and $B_j$ we considered are the ``multiport beam
splitters'' measurements described in
 \cite{multiport} and which have in previous numerical searches
yielded highly non local quantum correlations \cite{zeilig,durt}. These
measurements are
parametrized by $d$
phases $(\phi_{A_i}^1,\ldots \phi_{A_i}^d)$ and 
$(\phi_{B_j}^1,\ldots \phi_{B_j}^d)$ and involve the following steps: first each 
party acts with the phase $\phi_{A_i}(m)$ or $\phi_{B_j}(m)$ on the state 
$|m\rangle$, they then both carry out a discrete
Fourier transform. This brings the state $\Psi$ to:
\begin{eqnarray}
\Psi&=&\frac{1}{d^{3/2}}\sum_{k,l,m=0}^{d-1}\mbox{exp}
\biggl[i\biggl(\phi_{A_i}(m)-\phi_{B_j}(m)\nonumber \\
&&+\frac{2\pi}{d}m(k-l)\biggr)\biggr] |k\rangle_a|l\rangle_b
\end{eqnarray}
Alice then measures $|k\rangle_a$ and Bob $|l\rangle_b$. The quantum   
probabilities (\ref{qcid}) thus take the form 
\begin{eqnarray}\label{pqmsearch}
P_{kl}^{QM}(A_i,B_j)&=&\frac{1}{d^3}|\sum_{m=0}^{d-1} 
\mbox{exp}\Bigl[i\Bigl(\phi_{A_i}(m)-\phi_{B_j}(m) \nonumber \\ 
&&\quad +\frac{2\pi 
m}{d}(k-l)\Bigr)\Bigr]|^2  \nonumber \\ 
P^{QM}_k(A_i)&=&1/d \nonumber \\
P^{QM}_l(B_j)&=&1/d
\end{eqnarray}

The search for minimal $\eta_*^{\lambda=1}$ and $\eta_*^{\forall \lambda}$
then reduces to a non-linear optimization problem over Alice's and Bob's
phases. For this, we used the ``amoeba" search procedure with its starting point
fixed by the result of a randomized search algorithm.

Note that these searches are time-consuming. Indeed, the first part of the
computation, the solution to the linear problem, involves the optimization of
$(d+1)^{N_a+N_b}$ parameters, the classical probabilities $p_{\bf K L}$ (the
situation is even worse for $\eta_*^{\lambda}$, since the linear problem has to
be solved several times while performing a dichotomic search for $\eta_*^{\lambda}$).
Then when searching for the optimal measurements, the first part of the algorithm
has to be performed for each phase settings. This results in a rapid exponential
growth of the time needed to solve the entire problem with the dimension and the
number of settings involved. A second factor that complicates the search for 
optimal measurements is that, due to the relatively large number of 
parameters that the algorithm has to optimize, it can fail to find the 
global minimum and converge to a local minimum. This is one of the reasons why, 
as a first step, we restricted our searches to "multiport beam splitter" 
measurements since the number of parameters needed to describe them is much 
lesser than that for general Von Neumann measurements.  

Our results for setups our computers could handle in reasonable time 
are summarized in table I. In dimension 2, we also performed more general 
searches using von Neuman measurements but the results we obtained where the 
same as for the multiport beam splitters described above

\subsection{Optimal Bell inequalities}\label{secBI}

Upon finding the optimal quantum measurements and the corresponding
values of $\eta_{*}$, we have tried to find the Bell inequalities
which yield these threshold detector efficiencies. This is
essential to confirm analytically these numerical results but also in order for 
them to have practical significance, ie. to be possible to implement them 
in an experiment. 

To find these inequalities, we have used the approach developped in 
\cite{collins}. The first idea of this approach is to make use of the symmetries 
of the quantum probabilities and to search for Bell inequalitites which have the 
same symmetry. Thus for instance if $P(A_i=k, B_j=l) = P(A_i=k+m \mod d , B_j=l 
+ m \mod d)$ for all $m\in\{0,\ldots,d-1\}$, then it is useful to introduce the 
probabilities 
\begin{eqnarray} 
P(A_i = B_j + n) &=& \sum_{m=0}^{d-1} P(A_i=m,B_j=n + m \mod d) \nonumber\\ 
P(A_i \neq B_j + n) &=& \sum^{d-1}_{m=0 \atop l\neq n} P(A_i=m,B_j=l+m \mod d) 
\nonumber \\ \end{eqnarray} 
and to search for Bell inequalities written as linear combinations of the $P(A_i 
= B_j + n)$. This reduces considerably the number of Bell inequalities among 
which one must search in order to find the optimal one. The second idea is to 
search for the logical contradictions which force the Bell inequality to take a 
small value in the case of lhv theories. Thus the Bell inequality will contain 
terms with different weights, positive and negative, but the lhv theory cannot
satisfy all the relations with the large positive weights. Once we
had identified a candidate Bell inequality, we ran a computer program
that enumerated all the deterministic classical strategies and
computed the maximum value of the Bell inequality. The deterministic
classical strategies are those for which the probabilities
 $p_{K_1\ldots K_{N_a}L_1\ldots L_{N_b}}$ are equal either to $0$ or
 to $1$. In order to find the maximum classical value of a Bell
 expression, it suffices to consider them since the other strategies are 
obtained as convex combinations of the deterministic ones.

However when the number of settings, $N_a$ and $N_b$, and the
dimensionality $d$ increase, it becomes more and more difficult to
find the optimal Bell inequalities using the above analytical
approach. We therefore developped an alternative method based on the
numerical algorithm which is used to find the threshold detection
efficiency.

The idea of this numerical approach is based on the fact that the probabilities 
for which there exists a solution $p_{\bf KL}$ to eqs. (\ref{clprob},\ref{cond1},\ref{poscond}) form a convex
polytope whose vertices are the deterministic strategies. The facets of this 
polytope are hyperplanes of dimension $D-1$ where $D$ is the dimension of the 
space in which lies the polytope ($D$ is lower than the dimension $(d+1)^{N_a 
+N_b}$ of the total space of probabilities due to constraints such as the
normalizations conditions (\ref{cond1}) and the no-signalling conditions (\ref{nosignalling})). These
hyperplanes of dimension $D-1$ correspond to Bell inequalities. 

At the threshold $\eta_*$, the quantum probability $P^{QM}_{\lambda 
\eta_*}$ belongs to the boundary, i.e to one of the faces, of the polytope 
determined by eqs (\ref{clprob},\ref{cond1},\ref{poscond}). The solution 
$p_{\bf KL}^*$ to these equations at the threshold is computed by our algorithm 
and it corresponds to the convex combinations of deterministic strategies that 
reproduce the quantum correlations. From this solution it is then possible to 
construct a Bell inequality. Indeed, the face $F$ to which $P^{QM}_{\lambda 
\eta_*}$ belongs is the plane passing through the deterministic strategies 
involved in the convex combination $p_{\bf KL}^*$. Either, this face $F$ is a 
facet, i.e. an hyperplane of dimension $D-1$, or $F$ is of dimension lower 
than $D-1$. In the first case, the hyperplane $F$ correspond to the Bell 
inequality we are looking. In the second case, there is an infinity of 
hyperplanes of dimension $D-1$ passing by $F$, indeed every vector $\vec v$ 
belonging to the space orthogonal to the face $F$ determines such an hyperplane. 
To select one of these hyperplanes liyng outside the polytope, and thus 
corresponding effectively to a Bell inequality, we took as vector $\vec v$ 
the component normal to $F$ of the vector which connects the center of the 
polytope and the quantum probabilities when $\eta=1$: $P^{QM}_{\lambda \eta=1}$. 
Though this choice of $\vec v$ is arbitrary, it yields Bell inequalities which 
preserve the symmetry of the probabilities $P^{QM}$. 

As in the analytical method given above,  
we have verified by enumeration of the deterministic strategies 
that this hyperplane is indeed a Bell
inequality (ie. that it lies on one side of the polytope) and that it
yields the threshold detection efficiency $\eta_*$.

\section{Results}\label{secRE}

Our results are summarized in table I. We now describe them in more detail.

\subsection{Arbitrary dimension, two settings on each side ($N_a =
    N_b=2$).}\label{secRE2x2}

For dimensions up to 7, we found numerically that
$\eta_{*}^{\lambda=1}=\eta_{*}^{\forall \lambda}$. The optimal measurements 
we found are identical to those maximizing the  generalization of the CHSH
inequality to higher dimensional systems  \cite{collins}, thus confirming their 
optimality not only for the  resistance to noise but also for the resistance to 
inefficient  detectors. Our values of $\eta_*$ are identical to
those given in \cite{durt} where $\eta_{*}^{\lambda=1}$ has been calculated for 
these particular settings for $2\leq d \leq 16$.

We now derive a Bell inequality that reproduces analytically
these numerical results (which has also been derived by N. Gisin 
\cite{lambda}). Our Bell inequality is based on the generalization of 
the CHSH inequality obained in \cite{collins}. We recall the form of the Bell 
expression used in this inequality: \begin{eqnarray}
\label{chsh}
\lefteqn{I^{d,2\times 2}_{rr}=\sum_{k=0}^{[d/2]-1}\left(1-\frac{2k}{d-1}\right)} \nonumber \\
& \Bigl(+&[P(A_1=B_1+k)+P(B_1=A_2+k+1)\nonumber \\
&& +P(A_2=B_2+k)+P(B_2=A_1+k)]\nonumber \\
&-&[P(A_1=B_1-k-1)+P(B_1=A_2-k) \nonumber \\
&&+P(A_2=B_2-k-1)+P(B_2=A_1-k-1)]\Bigr). \nonumber \\
\end{eqnarray} 
For local theories, $I^{d,2\times 2}_{rr}\leq 2$ as shown in \cite{collins} 
where the value of $I^{d,2\times 2}_{rr}(QM)$ given by the optimal quantum 
measurements is also described.
In order to take into account ``no-result" outcomes we introduce the following
inequalities: 
\begin{equation} \label{chsheta}
I^{d,2\times 2}=I^{d,2\times 
2}_{rr}+\frac{1}{2}\sum_{i,j}P(A_i=\emptyset,B_j=\emptyset)\leq 2 
\end{equation}
Let us prove that the maximal allowed value of $I^{d,2\times 2}$
for local theories is $2$. To this end it suffices to enumerate all the 
deterministic strategies. First, if all the local variables correspond to a 
"result" outcome then $I^{d,2 \times 2}_{rr}\leq 2$ and $I^{d,2 \times 
2}_{\emptyset \emptyset}=\frac{1}{2}\sum_{i,j}P(A_i=\emptyset,B_j=\emptyset)=0$ 
so that $I^{d,2\times 2}\leq 2$; if one of the local variables is equal to 
$\emptyset$ then again $I^{d,2\times 2}_{rr}\leq 2$ (since the maximal weight of 
a probability in $I^{d,2\times 2}_{rr}$ is one and they are only two such 
probabilities different from zero) and $I^{d,2\times 2}_{\emptyset   
\emptyset}=0$; if there are two $\emptyset$ outcomes, then $I^{d,2\times 2}_{rr} 
\leq 1$ and $I^{d,2\times 2}_{\emptyset
\emptyset}\leq 1$; while if there are three or four $\emptyset$ then
$I^{d,2\times 2}_{rr}=0$ and
$I^{d,2\times 2}_{\emptyset \emptyset}\leq 2$.

Note that the inequality (\ref{chsheta}) obeys the condition 
$\sum_{i,j}c_{ij}^{\emptyset \emptyset}=c$, hence it will provide a bound on
$\eta_{*}^{\forall \lambda}$. Using eq. (\ref{etathr}), we obtain the value of 
$\eta_{*}^{\forall \lambda}$: \begin{equation} \label{etachsh} \eta_{*}^{\forall
\lambda}=\frac{4}{I^{d,2\times 2}_{rr}(QM)+2} \end{equation} 
Inserting the optimal values of $I^{d,2\times 2}_{rr}(QM)$ given in 
\cite{collins} this reproduces our numerical results and those of \cite{durt}. 
As an example, for dimension 3, $I^{3,2\times 2}_{rr}(QM)=2.873$ so that
$\eta_{*}^{\forall \lambda}=0.8209$. When $d\rightarrow\infty$,
(\ref{etachsh})
gives the limit $\eta_{*}^{\forall \lambda}=0.8049$.

\subsection{ 3 dimensions, $2\times 3$ settings.}

For three-dimensional systems, we found that adding
one setting to one of the party decreases both  $\eta_{*}^{\lambda
  = 1}$  and $\eta_{*}^{\forall
  \lambda}$ from $0.8209$ to $0.8182$ (In the case of $d=2$, it is necessary to 
take three settings on each side to get an improvement). The optimal settings 
involved are $\phi_{A_1}=(0,0,0)$, $\phi_{A_2}=(0,2\pi/3,0)$, 
$\phi_{B_1}=(0,\pi/3,0)$, $\phi_{B_2}=(0,2\pi/3,-\pi/3)$, 
$\phi_{B_3}=(0,-\pi/3,-\pi/3)$.

We have derived a Bell expression
associated to these measurements:
\begin{eqnarray}\label{2x3}
\lefteqn{I^{3,2\times 3}_{rr}=+[P(A_1=B_1)+P(A_1=B_2)+P(A_1=B_3)}\nonumber\\
&&+P(A_2=B_1+1)+P(A_2=B_2+2)+P(A_2=B_3)] \nonumber \\
&-&[P(A_1\neq B_1)+P(A_1\neq B_2)+P(A_1\neq B_3) \nonumber \\
&&+P(A_2\neq B_1+1)+P(A_2\neq B_2+2)+P(A_2\neq B_3)] \nonumber \\
\end{eqnarray}
The maximal value of $I^{3,2\times 3}_{rr}$ for classical theories is 2
since for any choice
of local variables 4 relations with a + can be satisfied but then two with a -
are also satisfied. For example we can satisfy the first four relations but this
implies $A_2=B_2+1$ and $A_2=B_3+1$ which gives 2 minus terms. The
maximal value of $I^{3,2\times 3}_{rr}$ for quantum mechanics is given for
the settings described above and is equal to $I^{3,2\times3}_{rr}(QM)=10/3$. To 
take into account detection inefficiencies consider the following inequality:
\begin{eqnarray}
\label{ineq2x3}
I^{3,2\times 3}=I^{3,2\times 3}_{rr}+I^{3,2\times 3}_{\emptyset
  r}+I^{3,2\times 3}_{\emptyset \emptyset} \leq 2
\end{eqnarray}
where
\begin{equation}
I^{3,2\times 3}_{\emptyset r}=-\frac{1}{3}\sum_{i,j}
P(A_i=\emptyset, B_j\neq \emptyset)
\end{equation}
and
\begin{equation}
I^{3,2\times 3}_{\emptyset \emptyset}=\frac{1}{3}\sum_{i,j} P(A_i=\emptyset,
B_j=\emptyset).
\end{equation}
($I_{r \emptyset}$ is taken equal to zero).
The principle used to show that $I^{3,2\times 3} \leq 2$, is the same as 
the one used to prove that $I^{d,2\times 2}\leq 2$. For example if
$A_1=\emptyset$ then $I^{3,2\times3}_{rr}\leq 3$, $I^{3,2\times 3}_{\emptyset 
r}=-1$ and $I^{3,2\times 3}_{\emptyset \emptyset}=0$ so that $I^{3,2\times 
3}\leq 3-1=2$. From (\ref{ineq2x3}) and the joint probabilities 
(\ref{pqmsearch}) for the optimal quantum measurements we deduce: 
\begin{equation}\label{2x3thr}   
\eta_{*}^{\forall
\lambda}=\frac{6}{\frac{10}{3}+4}=\frac{9}{11}\simeq 0.8182 
\end{equation} 
in agreement with our numerical result.

Note that in \cite{helle}, an inequality formally idendical to (\ref{2x3}) has 
been introduced. However, the measurement scenario involve two measurements on 
Alice's side and nine binary measurements on Bob's side. By grouping 
appropiately the outcomes, this measurements scenario can be associated to an 
inequality formally identical to (\ref{2x3}) for which the violation reaches 
$2\sqrt{3}$. According to (\ref{2x3thr}), this result in a detection efficiency 
threshold $\eta_{*}^{\forall \lambda}$ of $6/(2\sqrt{3}+4)\approx 0.8038$.

\subsection{3 settings for both parties}\label{secRE3x3}
For 3 settings per party, things become more surprising.  We have found
measurements that lower $\eta_{*}^{\lambda=1}$
and $\eta_{*}^{\forall \lambda}$ with respect to $2\times 2$ or $2\times 3$
settings. But contrary to the previous situations, $\eta_{*}^{\lambda=1}$ is
not equal to $\eta_{*}^{\forall \lambda}$, and the two optimal values are
obtained for two different sets of measurements. We present in this section the 
two Bell inequalities associated to each of these situations for the qubit case. 
Let us first begin with the inequality for $\eta_{*}^{\lambda=1}$: 
\begin{eqnarray}\label{3x3} \lefteqn{I^{2,3\times 3,\lambda}_{rr}=E(A_1,B_2)+E
(A_1,B_3)+E(A_2,B_1)} \nonumber \\
&&+E(A_3,B_1)-E(A_2,B_3)-E(A_3,B_2) \nonumber \\
&& -\frac{4}{3}P(A_1\neq
B_1)-\frac{4}{3}P(A_2\neq B_2)-\frac{4}{3}P(A_3\neq B_3)\leq 2 \nonumber \\
\end{eqnarray}
where
$E(A_i,B_j)=P(A_i=B_j)-P(A_i\neq B_j)$. As usually, the fact that $I^{2,3\times
3}_{rr}\leq 2$ follows from considering all deterministic classical 
strategies. The 
maximal quantum mechanical violation for this inequality is $3$ and is obtained 
by performing the same measurements on both sides $A_1=B_1$, $A_2=B_2$, 
$A_3=B_3$ defined by the following phases: $\phi_{A_1}=(0,0)$, 
$\phi_{A_2}=(0,\pi/3)$, $\phi_{A_3}=(0,-\pi/3)$. It is interesting to note that 
this inequality and these settings are related to those considered by Bell 
\cite{bell} and Wigner \cite{wigner} in the first works on quantum non-locality. 
But whereas in these works it was necessary to suppose that $A_i$ and $B_j$ are 
perfectly (anti-)correlated when $i=j$ in order to derive a contradiction with 
lhv theories, here imperfect correlations $P(A_i\neq B_i)>0$ can also lead to a 
contradiction since they are included in the Bell inequality.  

If we now consider "no-result" outcomes, we can use 
$I^{2,3\times 3,\lambda}_{rr}$ without adding extra
terms and the quantum correlations obtained from the optimal
measurements violate the
inequality if
\begin{equation}
\lambda\eta^2>\frac{2}{I^{2,3\times
3,\lambda}_{rr}(QM)}=\frac{2}{3}
\end{equation}
Taking $\lambda=1$, we obtain
$\eta_{*}^{\lambda=1}=\sqrt{2/3}\simeq 0.8165$. For smaller value of
$\lambda$, $\eta_{*}^{\lambda}$ increase until
$\eta_{*}^{\lambda}=16/19$ is
reached for $\lambda\simeq 0.9401$. At that point the contradiction with local
theories ceases to depend on the production rate $\lambda$. It is then 
advantageous to use the following inequality 
\begin{eqnarray}\label{3x3b}
\lefteqn{I^{2,3\times 3,\forall 
\lambda}_{rr}=\frac{2}{3}E(A_1,B_2)+\frac{4}{3}E
(A_1,B_3)+\frac{4}{3}E(A_2,B_1)} \nonumber \\
&&+\frac{2}{3}E(A_3,B_1)-\frac{4}{3}E(A_2,B_3)-\frac{2}{3}E(A_3,B_2) \nonumber 
\\ && -\frac{4}{3}P(A_1\neq
B_1)-\frac{4}{3}P(A_2\neq B_2)-\frac{4}{3}P(A_3\neq B_3)\leq 2 \nonumber \\
\end{eqnarray}
This inequality is similar to the former one (\ref{3x3}) but the symmetry 
between the $E(A_i,B_j)$ terms has been broken: half of the terms have an
additional weight of $1/3$ and the others of $-1/3$.
The total inequality involving "no-result" outcomes is
\begin{equation}
I^{2,3\times3,\forall \lambda}=I^{2,3\times 3,\forall \lambda}_{rr}+I_{\emptyset 
r}^{2,3\times 3,\forall \lambda}+I_{r \emptyset}^{2,3\times 3,\forall 
\lambda}+I^{2,3\times 3,\forall \lambda}_{\emptyset \emptyset}\leq 2
\end{equation}
The particular form of the terms $I_{\emptyset 
r}^{2,3\times 3,\forall \lambda}$, $I_{r \emptyset}^{2,3\times 3,\forall 
\lambda}$ and $I^{2,3\times 3,\forall \lambda}_{\emptyset \emptyset}$ is given 
in the Appendix. The important point is that 
$\sum_{i,j,k}(c_{ij}^{k \emptyset}+c_{i,j}^{\emptyset k})=-8/3$ and 
$\sum_{i,j}c_{ij}^{\emptyset \emptyset}=2$. From (\ref{etathr}), (\ref{belexpr}) and 
(\ref{pqmsearch}), we thus deduce \begin{equation} \eta_{*}^{\forall 
\lambda}=\frac{4+\frac{4}{3}}{I^{2,3\times 3,\forall 
\lambda}_{rr}(QM)+2+\frac{4}{3}} \end{equation}
The measurements that optimize the former inequality (\ref{3x3}) give the 
threshold $\eta_{*}^{\forall \lambda}=16/19$. However these measurements are not 
the optimal ones for (\ref{3x3b}). The optimal phase settings are given 
in the Appendix. Using these settings it follows that $ I^{2,3\times 3,\forall 
\lambda}_{rr}(QM)=3.157$ and $\eta_{*}^{\forall \lambda}\simeq 0.8217$. 

One may argue that the situation we have presented here is artificial 
and results from the fact that we failed to find the optimal inequality valid 
for all lambda which would otherwise have given a threshold 
$\eta_*^{\forall \lambda}=0.8165$ identical to the threshold 
$\eta_*^{\lambda=1}$. However, this cannot be the case since for $\lambda>1$ and 
$\eta>\eta_*^{\lambda=1}$ there exists a lhv model that reproduces the quantum 
correlations. This lhv model is simply given by the result of the first part 
of our algorithm described in \ref{secNS}. 

\subsection{More settings and more dimensions}
Our numerical algorithm has also yielded further improvements when the number
of settings increases or the dimension increases. These results are summarized in
Table I. For more details, see the Appendix.

\section{conclusion}
In summary we have obtained using both numerical and analytical techniques a 
large number of Bell inequalities and optimal quantum measurements that exhibit 
an enhanced resistance to detector inefficiency. This should be contrasted with 
the work (reported in \cite{zeilig, durt}) devoted to searching for Bell 
inequalities and measurements with increased resistance to noise. In this case 
only a single family has been found involving two settings on each side despite 
extensive numerical searches. Thus the structure of Bell inequalities resistant 
to inefficient detectors seems much richer. It would be interesting to 
understand the reason for such additional structure and clarify the origin of 
these inequalities.

It should be noted that for the Bell inequalities we have found, the amount 
by which the theshold detector efficiency $\eta_*$ decreases is very small, of 
the order of $4\%$.  This is tantalizing because we know that for sufficiently 
large dimension and sufficiently large number of settings, the detector efficiency
threshold decreases exponentially. To increase further the 
resistance to inefficent detector, it would perhaps be necessary to consider 
more general measurements than the one we considered in this work or use 
non-maximally entangled states (for instance, Eberhard has shown that for 
two-dimensional systems, the efficiency threshold $\eta_*$ can be lowered to 2/3 
using non-maximally entangled states \cite{eberhard}). There may thus be a Bell 
inequality of real practical importance for closing the detection loophole just 
behind the corner.

\begin{acknowledgments}
We would like to thank Nicolas Gisin for helpful discussions.
We acknowledges funding by the European Union under project EQUIP (IST-FET 
program). S.M. is a research associate of the Belgian
National Research Foundation. J.R. acknowledges support from the Belgian
FRIA.
\end{acknowledgments}

\appendix
\section{}

For completeness, we present here in details all the Bell inequalities and
optimal phase settings we have found. This includes also the results of Table I 
which have not been discussed in the text.

\begin{list}{$\bullet$}{\setlength{\leftmargin}{0.4cm}}
\item $\pmb{N_A=2,\:N_B=2, \: \forall \lambda}$

Bell inequality:
\begin{equation*}\begin{split}
I^{d,2\times 2}&=\sum_{k=0}^{[d/2]-1}\left(1-\frac{2k}{d-1}\right)  \\
& \Bigl(+[P(A_1=B_1+k)+P(B_1=A_2+k+1) \\
&+P(A_2=B_2+k)+P(B_2=A_1+k)] \\
&-[P(A_1=B_1-k-1)+P(B_1=A_2-k)  \\
&+P(A_2=B_2-k-1)+P(B_2=A_1-k-1)]\Bigr) \\
&+\frac{1}{2}\sum_{i,j=1}^{2}P(A_i=\emptyset,B_j=\emptyset) \leq 2
\end{split}\end{equation*}

Optimal phase settings:
\begin{equation*}\begin{array}{lll}
\phi_{A_1}(j)=0 & \phi_{A_2}(j)=\frac{\pi}{d}j  \\
\phi_{B_1}(j)=\frac{\pi}{2d}j & \phi_{B_2}(j)=-\frac{\pi}{2d}j
\end{array}\end{equation*}

Maximal violation:
\begin{equation*}
I^{d,2\times 2}(QM)=4d
\sum_{k=0}^{[d/2]-1}\left(1-\frac{2k}{d-1}\right)(q_k-q_{-k-1})
\end{equation*}
where $q_k=1/\left(2d^3\sin^2[\pi(k+1/4)/d]\right)$.

Detection threshold:
$\eta_*^{\forall \lambda}=\frac{4}{I^{d,2\times 2}(QM)+2}$

\item $\pmb{d=2,\:N_A=3,\:N_B=3,\:\lambda}$

Bell inequality:
\begin{equation*}\begin{split}
I^{2,3\times 3,\lambda}&=E(A_1,B_2)+E(A_1,B_3)\\
&+E(A_2,B_1)+E(A_3,B_1)-E(A_2,B_3)\\
&-E(A_3,B_2)-\frac{4}{3}P(A_1\neq B_1)\\
&-\frac{4}{3}P(A_2\neq B_2)-\frac{4}{3}P(A_3\neq B_3)\leq 2 \\
\end{split}\end{equation*}
where $E(A_i,B_j)=P(A_i=B_j)-P(A_i\neq B_j)$.

Optimal phase settings:
\begin{equation*}
\begin{array}{lll}
\phi_{A_1}=(0,0) & \phi_{A_2}=(0, \pi/3) & \phi_{A_3}=(0, -\pi/3)
\nonumber \\ 
\phi_{B_1}=(0,0)&\phi_{B_2}=(0, \pi/3)&\phi_{B_3}=(0, -\pi/3)
\end{array}
\end{equation*}

Maximal violation:
$I^{2,3\times 3,\lambda}(QM)=3$

Detection threshold:
$\eta_*^{\lambda}=\sqrt{\frac{2}{3\lambda}}$

\item $\pmb{d=2,\:N_A=3,\:N_B=3,\:\forall \lambda}$

Bell inequality:
\begin{equation*}\begin{split}
I&^{2,3\times 3,\forall\lambda}=\frac{2}{3}E(A_1,B_2)+\frac{4}{3}E
(A_1,B_3)\\
&+\frac{4}{3}E(A_2,B_1)+\frac{2}{3}E(A_3,B_1)-\frac{4}{3}E(A_2,B_3)\\
&-\frac{2}{3} E(A_3,B_2)-\frac{4}{3}P(A_1\neq B_1)\\
&-\frac{4}{3}P(A_2\neq B_2)-\frac{4}{3}P(A_3\neq B_3) \\
&-\frac{2}{3}F_{\emptyset}(A_1,B_2)-\frac{4}{3}F_{\emptyset}(A_2,B_3)-\frac{2}{3}F_{\emptyset}(
A_3,B_1) \\
&+\frac{2}{3}F_{\emptyset}(A_3,B_2)+\frac{4}{3}P(A_2=\emptyset,B_1\neq \emptyset) \\
&+\frac{4}{3}P(A_1\neq\emptyset, B_3=\emptyset)+\frac{4}{3}P(A_1=\emptyset,B_1=\emptyset)\\
&+\frac{4}{3}P(A_2=\emptyset,B_1=\emptyset)+\frac{4}{3}P(A_1=\emptyset,B_3=\emptyset) \leq 2 
\end{split}\end{equation*}
where $E(A_i,B_j)=P(A_i=B_j)-P(A_i\neq B_j)$ and $F_{\emptyset}(A_i,B_j)=P(A_i=\emptyset, 
B_j\neq\emptyset)+P(A_i\neq\emptyset,B_j=\emptyset)+P(A_i=\emptyset,B_j=\emptyset)$.

Optimal phase settings:
\begin{equation*}
\begin{array}{lll}
\phi_{A_1}=(0,0) & \phi_{A_2}=(0, 1.3934)\\
\phi_{A_3}=(0, -0.7558) \\      
\phi_{B_1}=(0,0.5525)&\phi_{B_2}=(0,1.3083) \\
\phi_{B_3}=(0,-0.8410) \\ 
\end{array} 
\end{equation*}

Maximal violation:
$I^{2,3\times 3, \forall \lambda}(QM)=3.157$
                                           
Detection threshold:
$\eta_*^{\forall \lambda}=0.8217$

\item $\pmb{d=2,\:N_A=3,\:N_B=4,\:\forall \lambda}$

Bell inequality:
\begin{equation*}\begin{split}
I&^{2,3\times 4, \forall \lambda}=-P(A_1\neq B_2)-P(A_1\neq B_3)-P(A_1\neq B_4) 
\\ &+P(A_2=B_1)+P(A_2=B_2)-P(A_2\neq B_3)\\
&+P(A_2\neq B_4)-P(A_3=B_1)+P(A_3=B_2)\\
&-P(A_3\neq B_2)+P(A_3\neq B_3)-P(A_3=B_4)\\
&+P(A_1\neq\emptyset,B_1=\emptyset)+P(A_2=\emptyset,B_1\neq\emptyset)\\
&-P(A_3\neq\emptyset,B_1=\emptyset)-P(A_1=\emptyset,B_2\neq\emptyset)\\
&+P(A_1=\emptyset,B_1=\emptyset)+P(A_2=\emptyset,B_2=\emptyset)\leq 2
\end{split}\end{equation*}

Optimal phase settings:
\begin{equation*}
\begin{array}{lll}
\phi_{A_1}=(0,0)&\phi_{A_2}=(0,0.7388)\\
\phi_{A_3}=(0,2.1334)\\
\phi_{B_1}=(0,-0.1347)&\phi_{B_2}=(0,1.2938)\\
\phi_{B_3}=(0,-0.0757)&\phi_{B_4}=(0 ,-1.0891)\\
\end{array}
\end{equation*}

Maximal violation:
$I^{2,3\times 4}(QM)=2.8683$

Detection threshold:
$\eta*^{\forall \lambda}=0.8216$

\item $\pmb{d=2,\:N_A=4,\:N_B=4,\:\forall \lambda}$

Bell inequality:
\begin{equation*}\begin{split}
I&^{2,4\times 4, \forall \lambda}=-P(A_1=B_1)+P(A_1\neq B_3)-P(A_2=B_1) 
\\ &-P(A_2=B_2)+P(A_2\neq B_4)+P(A_3\neq B_1)\\
&-P(A_3\neq B_2)-P(A_3\neq B_3)-P(A_4\neq B_1)\\
&-P(A_4=B_2)-P(A_4=B_3)+P(A_4\neq B_4)\\
&+P(A_1\neq\emptyset,B_4=\emptyset)-P(A_4\neq\emptyset,B_1=\emptyset)\\
&+P(A_1=\emptyset,B_1=\emptyset)+P(A_1=\emptyset,B_4=\emptyset)\leq 2
\end{split}\end{equation*}

Optimal phase settings:
\begin{equation*}
\begin{array}{lll}
\phi_{A_1}=(0,0)&\phi_{A_2}=(0,0.0958)\\
\phi_{A_3}=(0,2.1856)&\phi_{A_4}=(0,4.5944)\\
\phi_{B_1}=(0,4.0339)&\phi_{B_2}=(0,3.3011)\\
\phi_{B_3}=(0,2.2493)&\phi_{B_4}=(0, 2.3454)\\
\end{array}
\end{equation*}

Maximal violation:
$I^{2,4\times 4}(QM)=2.8697$

Detection threshold:
$\eta_*^{\forall \lambda}=0.8214$
\pagebreak
\item $\pmb{d=3,\:N_A=2,\:N_B=3, \: \forall \lambda}$

Bell inequality:
\begin{equation*}\begin{split}
I&^{3,2\times 3,\lambda}=+[P(A_1=B_1)+P(A_1=B_2)+P(A_1=B_3)\\
&+P(A_2=B_1+1)+P(A_2=B_2+2)+P(A_2=B_3)]  \\
&-[P(A_1\neq B_1)+P(A_1\neq B_2)+P(A_1\neq B_3)  \\
&+P(A_2\neq B_1+1)+P(A_2\neq B_2+2)+P(A_2\neq B_3)] \\
&-\frac{1}{3}\sum_{i,j} P(A_i=\emptyset, B_j\neq
\emptyset)  \\
&+\frac{1}{3}\sum_{i,j} P(A_i=\emptyset,B_j=\emptyset) \leq 2
\end{split}\end{equation*}

Optimal phase settings:
\begin{equation*}
\begin{array}{ll}
\phi_{A_1}=(0,0,0)&\phi_{A_2}=(0, 2\pi/3,0) \\
\phi_{B_1}=(0,\pi/3,0)&\phi_{B_2}=(0, 2\pi/3,-\pi/3) \\
\phi_{B_3}=(0,-\pi/3,-\pi/3) \\ 
\end{array}
\end{equation*}

Maximal violation:
$I^{3,2\times 3}(QM)=\frac{10}{3}$

Detection threshold:
$\eta_*^{\forall \lambda}=\frac{9}{11}\simeq 0.8182$

\item $\pmb{d=3,\: N_A=3,\:N_B=3,\:\lambda}$

Bell inequality:
\begin{equation*}\begin{split}
I&^{3,3\times 3,\lambda}=E_1(A_1,B_2)+E_2(A_1,B_3)\\
&+E_2(A_2,B_1)-E_2(A2,B_3)+E_1(A_3,B_1)\\
&-E_1( A _3,B_2)-P(A_1\neq B_1)\\
&-P(A_2\neq B_2)-P(A_3 \neq B_3) \leq 2
\end{split}\end{equation*}

Optimal phase settings:
\begin{equation*}
\begin{array}{ll}
\phi_{A_1}=(0,0,0)&\phi_{A_2}=(0,2\pi/9,4\pi/9)\\
\phi_{A_3}=(0,-2\pi/9,-4\pi/9) \nonumber \\ 
\phi_{B_1}=(0,0,0)&\phi_{B_2}=(0,2\pi/9,4\pi/9)\\
\phi_{B_3}=(0,-2\pi/9,-4 \pi/9 )  
\end{array}
\end{equation*}
          
Maximal violation:
$I^{3,3\times 3}(QM)=3.0642$

Detection threshold:
$\eta_*^{\lambda}=\frac{2}{3.0642\lambda}$
\pagebreak
\item $\pmb{d=3,\:N_A=3,\: N_B=3,\:\forall \lambda}$

Bell inequality:
\begin{equation*}\begin{split}
I&^{3,3\times 3, \forall \lambda}=-\frac{5}{3}P(A_1=B_1)-\frac{4}{3}P(A_1=B_1+2)\\ 
&+P(A_1= B_2)+{5\over 3}P(A_1=B_2+1)-{5\over 3}P(A_1=B_3)\\
&-P(A_1= B_3+2)+\frac{5}{3}P(A_2=B_1)-2P(A_2=B_1+1)\\ 
&-{5 \over 3}P(A_2= B_2)+2P(A_2=B_2+1)-P(A_2=B_3+1)\\
&-{5 \over 3}P(A_2= B_3+2)-\frac{11}{3}P(A_3=B_1)-2P(A_3=B_1+2)\\ 
&+{2 \over 3}P(A_3= B_2)+2P(A_3=B_2+1)+{5\over 3}P(A_3=B_3)\\
&+P(A_3= B_3+2)+{5\over 3}P(A_1\neq\emptyset,B_1=\emptyset)\\
&-{5\over 3}P(A_2\neq\emptyset,B_1=\emptyset)-2P(A_3\neq\emptyset,B_1=\emptyset)\\
&+2P(A_1\neq\emptyset,B_2=\emptyset)+{5\over 3}P(A_1=\emptyset,B_1=\emptyset)\\
&+2P(A_1=\emptyset,B_2=\emptyset)\leq 11/3
\end{split}\end{equation*}

Optimal phase settings:
\begin{equation*}\begin{array}{ll}
\phi_{A_1}=(0,0,0) & \phi_{A_2}=(0, 1.4376,2.8753)\\
\phi_{A_3}=(0,0.5063,1.0125) \\
\phi_{B_1}=(0,2.0452,4.0904)& \phi_{B_2}=(0,2.9758,-0.3315) \\ 
\phi_{B_3}=(0,1.3839, 2.7678) \\
\end{array}\end{equation*}

Maximal violation:
$I^{3,3\times 3}(QM)=5.3358$

Detection threshold:
$\eta_*^{\forall \lambda}=0.8146$

\item $\pmb{d=4,\:N_A=2,\:N_B=3,\:\forall \lambda}$

Bell inequality:
\begin{equation*}\begin{split}
I&^{4,2\times 3, \forall \lambda}=P(A_1=B_1+1)+2P(A_1=B_1+2)\\ 
&+2P(A_1= B_2)+P(A_1=B_2+1)+2P(A_1=B_3)\\ 
&+2P(A_2=B_1+1)+P(A_2=B_1+2)+P(A_2=B_2)\\ 
&+2P(A_2=B_2+1)+2P(A_2=B_3+2)\\
&+{4\over 3}\sum_iP(A_i=\emptyset,B_1\neq\emptyset)+ {1\over 
3}\sum_iP(A_i=\emptyset,B_2\neq\emptyset)\\ 
&+{1\over 3}\sum_iP(A_i=\emptyset,B_3\neq\emptyset)+{5\over 
3}\sum_iP(A_1\neq\emptyset,B_1=\emptyset)\\ 
&+{1\over 3}\sum_iP(A_2\neq\emptyset,B_1=\emptyset)+{8\over 3}P(A_1=\emptyset,B_1=\emptyset)\\
&+{5\over 3}P(A_1=\emptyset,B_2=\emptyset)+{5\over 3}P(A_1=\emptyset,B_3=\emptyset)\\
&+{4\over 3}P(A_2=\emptyset,B_1=\emptyset)+{1\over 3}P(A_2=\emptyset,B_2=\emptyset)\\
&+{1\over 3}P(A_2=\emptyset,B_3=\emptyset)\leq 8 \\ 
\end{split}\end{equation*}

Optimal phase settings:
\begin{equation*}\begin{array}{l}
\phi_{A_1}=(0,0,0,0)\\
\phi_{A_2}=(0,-1.1397,2.0019,3.1416)\\
\phi_{B_1}=(0,1.7863,-0.5698,2.3562)\\
\phi_{B_2}=(0,0.2155,5.7133,0.7854)\\
\phi_{B_3}=(0,1.0009,1.0009,0)
\end{array}\end{equation*}

Maximal violation:
$I^{4,2\times 3}(QM)=9.4142$

Detection threshold:
$\eta_*^{\forall \lambda}=0.8093$

\item $\pmb{d=4,\:N_A=3,\:N_B=3,\:\forall \lambda}$

Bell inequality:
\begin{equation*}\begin{split}
I&^{4,3\times 3, \forall \lambda}=-P(A_1=B_1+2)+P(A_1=B_1+3)\\ 
&+2P(A_1= B_2+1)-P(A_1=B_2+2)-P(A_1=B_3)\\
&-3P(A_1=B_3+1)-2P(A_1=B_3+2)-P(A_2=B_1)\\
&+P(A_2=B_1+1)-P(A_2=B_2+1)+P(A_2=B_2+2)\\
&+2P(A_2=B_3+3)+2P(A_3=B_1+1)+P(A_3=B_2)\\
&-2P(A_3=B_2+2)-P(A_3=B_2+3)+2P(A_3=B_3)\\
&+P(A_3=B_3+2)+\sum_iP(A_i=\emptyset,B_1\neq\emptyset)\\
&+P(A_1\neq\emptyset,B_1=\emptyset)+P(A_1\neq\emptyset,B_2=\emptyset)\\
&-P(A_1=\emptyset,B_3\neq \emptyset)+P(A_3=\emptyset,B_3\neq\emptyset)\\
&+P(A_3\neq\emptyset,B_3=\emptyset)+2P(A_1=\emptyset,B_1=\emptyset)\\
&+P(A_1=\emptyset,B_2=\emptyset)+P(A_2=\emptyset,B_1=\emptyset)\\
&+P(A_3=\emptyset,B_1=\emptyset)+P(A_3=\emptyset,B_3=\emptyset)\leq 6\\ 
\end{split}\end{equation*}

Optimal phase settings:
\begin{equation*}\begin{array}{l}
\phi_{A_1}=(0,0,0,0)\\
\phi_{A_2}=(0,-1.2238,-1.1546,3.9048)\\
\phi_{A_3}=(0,3.1572,3.8330,0.7070)\\
\phi_{B_1}=(0,-0.9042,1.7066,0.8025)\\
\phi_{B_2}=(0,2.5844,3.6937,-0.0051)\\
\phi_{B_3}=(0,4.1396,3.0022,7.1419)
\end{array}\end{equation*}

Maximal violation:
$I^{4,3\times 3}(QM)=7.5576$

Detection threshold:
$\eta_*^{\forall \lambda}=0.7939$

\end{list}


\begin{thebibliography}{99}

\bibitem{bell} J.S. Bell, Physics {\bf 1}, 195 (1964)
\bibitem{weihs} G. Weihs, T. Jennewein, C. Simon, H. Weinfurter and 
A. Zeilinger, Phys. Rev. Lett. {\bf 81}, 5039 (1998) 
\bibitem{rowe} M. A. Rowe, D. Kielpinski, V. Meyer, C. A. Sackett, W. M. Itano, 
C. Monroe and D. J. Wineland, Nature {\bf 409}, 791 (2001) 
\bibitem{zeilig} D. Kaszlikowski, P. Gnaci\'nski, M. \.Zukowski, W. 
Miklaszewski and A. Zeilinger, Phys. Rev. Lett. {\bf 85}, 4418 (2001) 
\bibitem{durt} T. Durt, D. Kaszlikowski and M. \.Zukowski, Phys. Rev. A {\bf 
64}, 024101 (2001) 
\bibitem{collins} D. Collins, N. Gisin, N. Linden, S. Massar and S. Popescu, 
Phys. Rev. Lett. {\bf 88} 040404 (2002)
\bibitem{massar} S. Massar, Phys. Rev. A {\bf 65} 032121 (2002)
\bibitem{gisin} N. Gisin and B. Gisin, Phys. Lett. A {\bf 260} 323 (1999)
\bibitem{lhv} E. Santos, Phys. Rev. A {\bf 46}, 3646 (1992) 
\bibitem{inprep} S. Massar and S. Pironio, in preparation. 
\bibitem{chsh} J. F. Clauser, M. A. Horne, A. Shimony and R. A. Holt, Phys. Rev. 
Lett {\bf 23}, 880 (1969) 
\bibitem{multiport} M. \.Zukowski, A. Zeilinger and M. A. Horne, Phys. Rev. A 
{\bf 55}, 2564 (1997)
\bibitem{lambda} N. Gisin, private   communication.
\bibitem{zuk} M. \.Zukowski, D. Kaszlikowski, A. Baturo and J. Larsson, 
quant-ph/9910058
\bibitem{kaszli} D. Kaszlikowski, L. C. Kwek, J. Chen, M. \.Zukowski and C. H. Oh, Phys. Rev. A {\bf 65} 032118 (2002)
\bibitem{helle} H. Bechmann-Pasquinucci and N. Gisin, quant-ph/0204122
\bibitem{wigner} E. P. Wigner, Am. J. Phys. {\bf 38}, 1005 (1970).
\bibitem{eberhard} P. H. Eberhard, Phys. Rev. A {\bf 47}, R747 (1993) 
\end{thebibliography}
\end{document}